\documentclass[10pt]{article}
\usepackage{amsmath}
\usepackage{mathrsfs}
\usepackage{graphicx}
\usepackage{amsfonts}

\begin{document}

\vspace*{-.6in} \thispagestyle{empty}

\baselineskip = 18pt

\vspace{.5in}

\begin{center}
{\Large \textbf{Possible theoretical limits on holographic
quintessence from weak gravity conjecture}}
\end{center}

\vspace*{0.1cm}

\begin{center}
Yin-Zhe Ma$^{a,b}$ and Xin Zhang$^{a}$

\emph{$^{a}$Kavli Institute for Theoretical Physics China, Institute of
Theoretical Physics, Chinese Academy of Sciences (KITPC/ITP-CAS), P.O.Box
2735, Beijing 100080, People's Republic of China\\[0pt]
$^{b}$Graduate School of the Chinese Academy of Sciences, Beijing 100039,
People's Republic of China}

%\emph{$^{c}$ Kavli Institute for Theoretical Physics China }

\vspace{.1in}
\end{center}

\vspace{.2in}

\begin{center}
\textbf{Abstract}
\end{center}

\begin{quotation}
The holographic dark energy model is one of the important ways for
dealing with the dark energy problems in the quantum gravity
framework. In this model, the dimensionless parameter $c$ plays an
essential role in determining the evolution of the holographic dark
energy. In particular, the holographic dark energy with $c\geq 1$
can be effectively described by a quintessence scalar-field.
However, according to the requirement of the weak gravity conjecture
the variation of the quintessence scalar-field should be less than
the Planck mass, which would give theoretic constraints on the
parameters $c$ and $\Omega _{\mathrm{m0}}$. Therefore, we get the
possible theoretical limits on the parameter $c$ for the holographic
quintessence model.
\end{quotation}

\vfil \centerline{}

\newpage

\addtocontents{toc}{\protect\setcounter{tocdepth}{2}} \pagenumbering{arabic}

\vspace{0.5in}
%%%%%%%%%%%%%%%%%%%%%%%%%%%%%%%%%%%%%%%%%%%%%%%%%%%%%%%%%%%%%%%%%%%%%%%%%%%%%%%%%%%%%%%%%%%%%%%%%
%%%%%%%%%%%%%%%%%%%%%%%%%%%%%%%%%%%%%%%%%%%%%%%%%%%%%%%%%%%%%%%%%%%%%%%%%%%%%%%%%%%%%%%%%%%%%

It has been realized firmly that our universe is experiencing an
accelerated expansion at the present time, through the astronomical
observations, such as observations of large scale structure (LSS)
\cite{LSS}, searches for type Ia supernovae (SNIa) \cite{SN}, and
measurements of the cosmic microwave background (CMB) anisotropy
\cite{CMB}. The acceleration of the universe strongly indicates the
existence of a mysterious exotic matter, namely the dark energy,
which has large enough negative pressure and has been a dominative
power of the universe (for reviews see e.g. \cite{cc}). The combined
analysis of observational data suggests that the universe is
spatially flat, and consists of approximately $70\%$ dark energy,
$30\%$ dust matter (cold dark matter plus baryons), and negligible
radiation. Although it can be affirmed that the ultimate destiny of
our universe is determined by the feature of dark energy, we still
know little about the nature of dark energy. However, one still can
propose some candidates to interpret or describe its various
properties. The most simple yet indispensable theoretical candidate
for dark energy is the Einstein's cosmological constant $\lambda $
(vacuum energy) \cite{Einstein:1917} which has the equation of state
$w_{\lambda }=-1$. However, as is well known, the cosmological
constant scenario is always plagued with the two famous cosmological
constant problems regarding why $\rho _{\lambda }$ is much smaller
than many known contributions to it and why it is comparable to the
energy density of matter today .

Another candidate for dark energy is the energy density associated
with dynamical scalar-field, a slowly varying, spatially homogeneous
component. An example of scalar-field dark energy is the so-called
quintessence \cite{quintessence}, a scalar field $\phi $ slowly
evolving down its potential $V(\phi )$. Provided that the evolution
of the field is slow enough, the kinetic energy density is less than
the potential energy density, giving rise to the negative pressure
responsible to the cosmic acceleration. So far a wide variety of
scalar-field dark energy models have been proposed. Besides
quintessence, these also include phantom \cite{phantom}, $K$-essence
\cite{kessence}, tachyon \cite{tachyon}, ghost condensate
\cite{ghost} and quintom \cite{quintom} amongst many. However, we
should note that the mainstream viewpoint regards the scalar-field
dark energy models as an low-energy effective description of the
underlying theory of dark energy. In addition, other proposals on
dark energy include interacting dark energy models \cite{intde},
variable cosmological constant models \cite{vcc}, braneworld models
\cite{brane}, and
Chaplygin gas models \cite{cg}, etc. %One should realize, nevertheless, that
%almost all these models are settled at the phenomenological level, lacking
%theoretical roots.

Theoretical physicists have made lots of efforts trying to resolve
the cosmological constant problems, but all these efforts seem to be
unsuccessful. Of course the theoretical considerations have made
some progress and are still in process. In recent years, many string
theorists have devoted to shedding light on the cosmological
constant or dark energy problems within the string theory framework.
The famous Kachru-Kallosh-Linde-Trivedi (KKLT) model \cite{kklt} is
a typical example, which tries to construct metastable de Sitter
vacua in the light of type IIB string theory. Furthermore, string
landscape idea \cite{landscape} has been proposed for shedding light
on the cosmological constant problems based upon the anthropic
principle and multiverse speculation. Another way of endeavoring to
probe the nature of dark energy within the fundamental theory
framework originates from some considerations of the features of the
quantum gravity theory. It is generally believed by theorists that
we can not entirely understand the nature of dark energy before a
complete theory of quantum gravity is available. However, although
we are lacking a quantum gravity theory today, we still can make
some attempts to probe the nature of dark energy according to some
principles of quantum gravity. The holographic dark energy model
\cite{Li:2004rb} is just an appropriate example,
which is constructed in the light of the holographic principle \cite%
{holoprin} of quantum gravity theory. That is to say, the
holographic dark energy model possesses some significant features of
the underlying theory of dark energy.

According to the holographic principle, the number of degrees of
freedom for a system within a finite region should be finite and
bounded roughly by the area of its boundary. In the cosmological
context, the holographic principle will set an upper bound on the
entropy of the universe. Motivated by the Bekenstein entropy bound,
it seems plausible to require that for an
effective quantum field theory in a box of size $L$ with UV cutoff $\Lambda $%
, the total entropy should satisfy $S=L^{3}\Lambda ^{3}\leq S_{BH}\equiv \pi
M_{\mathrm{pl}}^{2}L^{2}$, where $S_{BH}$ is the entropy of a black hole
with the same size $L$. However, Cohen et al. \cite{Cohen:1998zx} pointed
out that to saturate this inequality some states with Schwartzschild radius
much larger than the box size have to be counted in. As a result, a more
restrictive bound, the energy bound, has been proposed to constrain the
degrees of freedom of the system, requiring that the total energy of a
system with size $L$ should not exceed the mass of a black hole with the
same size, namely, $L^{3}\Lambda ^{4}=L^{3}\rho _{\Lambda }\leq LM_{\mathrm{%
pl}}^{2}$. This means that the maximum entropy is in order of $S_{BH}^{3/4}$%
. When we take the whole universe into account, the vacuum energy
related to this holographic principle is viewed as dark energy,
usually dubbed ``holographic dark energy''. The largest IR cut-off
$L$ is chosen by saturating the inequality so that we get the
holographic dark energy density
\begin{equation}
\rho _{\Lambda }=3c^{2}M_{\mathrm{pl}}^{2}L^{-2}~,  \label{de}
\end{equation}%
where $c$ is a numerical constant (note that $c>0$ is assumed), and
as usual $M_{\mathrm{pl}}$ is the reduced Planck mass. It has been
conjectured by Li \cite{Li:2004rb} that the IR cutoff $L$ should be
given by the future event horizon of the universe
\begin{equation}
R_{\mathrm{eh}}(a)=a\int\limits_{t}^{\infty }{\frac{dt^{\prime }}{%
a(t^{\prime })}}=a\int\limits_{a}^{\infty }{\frac{da^{\prime }}{Ha^{\prime 2}%
}}~.  \label{eh}
\end{equation}%
Such a holographic dark energy looks reasonable, since it may
simultaneously provide natural
solutions to both dark energy problems as demonstrated in Ref. \cite%
{Li:2004rb}. The holographic dark energy model has been tested and
constrained by various astronomical observations
\cite{obshuang,GWZ,obszx1,obszx2,obsnew,obswei}. For other extensive
studies on the holographic dark energy, see e.g. Refs.
\cite{holoext1,holoext2}.

The holographic dark energy scenario reveals the dynamical nature of
the vacuum energy. When taking the holographic principle into
account, the vacuum energy density will evolve dynamically. Though
the underlying theory of dark energy is unavailable presently, we
can, nevertheless, speculate on the underlying theory of dark energy
by taking some principles of quantum gravity into account. The
holographic dark energy model is no doubt a tentative in this way.
Now, we are interested in that if we assume the holographic vacuum
energy scenario as the underlying theory of dark energy, how the
low-energy effective scalar-field model can be used to describe it.
In this direction, some work has been done. The holographic versions
of scalar-field models, such as
quintessence, tachyon, and quintom, have been constructed \cite%
{holoquin,holotachy,hologhost}. In this paper, we focus on the
canonical scalar-field description of the holographic dark energy,
namely the ``holographic quintessence'' \cite{holoquin}.

It is generally believed that string theory is the most promising
consistent theory of quantum gravity. By means of the KKLT mechanism
\cite{kklt} (see also \cite{fluxcmpctf}), a vast number of
meta-stable de Sitter vacua can be constructed through the flux
compactification on a Calabi-Yau manifold. These string vacua can be
described by the low-energy effective theories. However, recently,
it was realized that the vast series of semiclassically consistent
field theories are actually inconsistent. These actually
inconsistent effective field theories are viewed as located in the
so-called ``swampland'' \cite{swampland}. The self-consistent
landscape is surrounded by the swampland.

Undoubtedly, it is an important mission to distinguish the landscape
and the swampland. Vafa has proposed some criterion to the
consistent effective field theories \cite{swampland}. Furthermore,
recently, it was conjectured by Arkani-Hamed et al. \cite{weakgrav}
that the gravity is the weakest force, which helps to rule out those
effective field theories in the swampland. This conjecture is
supported by string theory and some evidence involving black holes
and symmetries \cite{weakgrav} (for the other arguments in string
theory to support this conjecture see also \cite{Ooguri:2006in}).
Arkani-Hamed et al. pointed out \cite{weakgrav} that when
considering the quantum gravity, the gravity and other gauge forces
should not be treated separately. For example, in four dimensions a
new intrinsic UV cutoff for the U(1) gauge theory, $\Lambda=g M_{\rm
pl}$, is suggested, where $g$ is the gauge coupling \cite{weakgrav}.
This conjecture was generalized to asymptotic dS/AdS background
\cite{hls}. In \cite{hls}, the weak gravity conjecture together with
the requirement that the IR cutoff should be smaller than the UV
cutoff leads to an upper bound for the cosmological constant. In
addition, for the inflationary cosmology, the application of the
weak gravity conjecture shows that the chaotic inflation model is in
the swampland \cite{Huangchaotic}. This conjecture even implies that
the eternal inflation may not be achieved \cite{HLW}. Furthermore,
Huang conjectured \cite{Huang:2007qz} that the variation of the
inflaton should be smaller than the Planck scale $M_{\rm pl}$, and
this can make stringent constraint on the spectral index.

Naturally, the weak gravity conjecture can also be applied to the
dark energy problem. This suggests that the variation of the
quintessence field value $\phi $ should be less than
$M_{\mathrm{pl}}$ \cite{Huang:2007mv}. This criterion may give
important theoretical constraints on the equation-of-state parameter
of quintessence models, and some of these constraints are even
stringent than those of the present experiments \cite{Huang:2007mv}.
In this paper we shall investigate the possible theoretical
constraints on the parameters of the holographic quintessence from
the weak gravity conjecture.

First, we briefly review the holographic dark energy model. Since
the spatial flatness is motivated by theoretical considerations
(such as the inflationary theory) and astronomical observations, we
consider a spatially flat universe filled with matter component
$\rho _{\mathrm{m}}$ (including both baryon matter and cold dark
matter) and holographic dark energy component $\rho _{\Lambda }$,
thus the Friedmann equation reads
\begin{equation}
3M_{\mathrm{pl}}^{2}H^{2}=\rho _{\mathrm{m}}+\rho _{\Lambda }~,
\label{friedmann}
\end{equation}%
or equivalently,
\begin{equation}
E(z)\equiv {\frac{H(z)}{H_{0}}}=\left( \frac{\Omega _{\mathrm{m0}}(1+z)^{3}}{%
1-\Omega _{\Lambda }}\right) ^{1/2},  \label{Ez}
\end{equation}%
where $z=(1/a)-1$ is the redshift of the universe. Combining the
definition
of the holographic dark energy (\ref{de}) and the definition of the future event horizon (\ref%
{eh}), we derive
\begin{equation}
\int_{a}^{\infty }{\frac{d\ln a^{\prime }}{Ha^{\prime }}}={\frac{c}{Ha\sqrt{%
\Omega _{\Lambda }}}}~.  \label{rh}
\end{equation}%
The Friedmann equation (\ref{Ez}) implies
\begin{equation}
{\frac{1}{Ha}}=\sqrt{a(1-\Omega _{\Lambda })}{\frac{1}{H_{0}\sqrt{\Omega _{%
\mathrm{m0}}}}}~.  \label{fri}
\end{equation}%
Substituting (\ref{fri}) into (\ref{rh}), one obtains the following equation
\begin{equation}
\int_{x}^{\infty }e^{x^{\prime }/2}\sqrt{1-\Omega _{\Lambda }}dx^{\prime
}=ce^{x/2}\sqrt{{\frac{1}{\Omega _{\Lambda }}}-1}~,
\end{equation}%
where $x=\ln a$. Then taking derivative with respect to $x$ in both sides of
the above relation, we easily get the dynamics satisfied by the dark energy,
i.e. the differential equation about the fractional density of dark energy,
\begin{equation}
\Omega _{\Lambda }^{^{\prime }}=-(1+z)^{-1}\Omega _{\Lambda }(1-\Omega
_{\Lambda })\left( 1+{\frac{2}{c}}\sqrt{\Omega _{\Lambda }}\right) ,
\label{deq}
\end{equation}%
where the prime denotes the derivative with respect to the redshift
$z$. This equation describes the behavior of the holographic dark
energy completely, and it can be solved exactly
\cite{Li:2004rb,obshuang}. From the energy conservation equation of
the dark energy, the equation of state of the dark energy can be
given by \cite{obshuang}
\begin{equation}
w_{\Lambda }=-1-{\frac{1}{3}}{\frac{d\ln \rho _{\Lambda }}{d\ln a}}=-{\frac{1%
}{3}}\left( 1+{\frac{2}{c}}\sqrt{\Omega _{\Lambda }}\right) ~.  \label{w}
\end{equation}%
Note that the formula $\rho _{\Lambda }={\frac{\Omega _{\Lambda }}{1-\Omega
_{\Lambda }}}\rho _{\mathrm{m0}}a^{-3}$ and the differential equation of $%
\Omega _{\Lambda }$ (\ref{deq}) are used in the second equal sign.

The property of the holographic dark energy is mainly governed by
the numerical parameter $c$. From Eq. (\ref{w}), it can be easily
found that the evolution of the
equation of state satisfies $-(1+2/c)/3\leq w_{\Lambda }\leq -1/3$ due to $%
0\leq \Omega _{\Lambda }\leq 1$. Thus, the parameter $c$ plays a
significant role in the holographic evolution of the universe. When
$c<1$, the
holographic evolution will make the equation of state cross $w=-1$ (from $%
w>-1$ evolves to $w<-1$); when $c\geq 1$, the equation of state will evolve
in the region of $-1\leq w\leq -1/3$.

Next, let us consider the quintessence scalar-field model. The quintessence
scalar field $\phi $ evolves in its potential $V(\phi )$ and rolls towards
its minimum of the potential, according to the Klein-Gordon equation $\ddot{%
\phi}+3H\dot{\phi}=-dV/d\phi $. The slope of the potential drives the rate
of evolution while the cosmic expansion damps this evolution through the
Hubble parameter $H$. The energy density and pressure are $\rho _{\phi }=%
\dot{\phi}^{2}/2+V$, $p_{\phi }=\dot{\phi}^{2}/2-V$, so that the
equation of state of quintessence $w_{\phi }=p_{\phi }/\rho _{\phi
}$ evolves in a region of $-1<w_{\phi }<1$. Usually, in order to
make the universe's expansion accelerate, $w_{\phi }$ should be
required less than $-1/3$. Nevertheless, it can be seen clearly that
the quintessence scalar field can not realize the equation of state
crossing $-1$. Therefore, only the holographic dark energy in cases
of $c\geq 1$ can be described by the quintessence
\cite{holoquin}.\footnote{Apparently, the quintessence model is
consistent with the second law of thermodynamics. In the holographic
dark energy model, the entropy of the whole system is described by
$S=\pi M_{\rm pl}^2R_{\rm eh}^2$. To satisfy the second law of
thermodynamics, one requires that $\dot{R_{\rm eh}}\geq 0$, which
leads to $c\geq\sqrt{\Omega_{\Lambda}}$ (for the general case in
non-flat space, see \cite{GWZ}). For the quintessence model, $w\geq
-1$, this together with Eq. (\ref{w}) also leads to
$c\geq\sqrt{\Omega_{\Lambda}}$. Furthermore, since the maximum of
$\Omega_\Lambda$ is 1, we thus obtain the condition $c\geq 1$ for
the quintessence-like behavior realization of the holographic dark
energy. }

In fact, in the holographic scenario, the value of $c$ should be
determined by cosmological observations. However, current
observational data are not precise enough to determine the value of
$c$ very accurately. An analysis of the latest observational data,
including the gold sample of 182 SNIa, the CMB shift parameter given
by the 3-year WMAP observations, and the baryon acoustic oscillation
(BAO) measurement from the Sloan Digital Sky Survey (SDSS), shows
that the possibilities of $c>1$ and $c<1$ both exist and their
likelihoods are almost equal within 3 sigma error range
\cite{obsnew}. Therefore, neither quintessence feature nor quintom
one can be ruled out by observational data presently available. In
\cite{obsnew}, the fit values for the model parameters with
1-$\sigma$ errors are $c=0.91^{+0.26}_{-0.18}$ and $\Omega_{\rm m0}=
0.29\pm 0.03$ with $\chi_{\rm min}=158.97$. Clearly, the range of
$c$ in the 1-$\sigma$ error, $0.73< c<1.17$, is not capable of
ruling out the probability of $c> 1$; this conclusion is somewhat
different from those derived from previous investigations using
earlier data. In previous work, for instance \cite{obszx1,obszx2},
the 1-$\sigma$ range of $c$ obtained can basically exclude the
probability of $c> 1$ giving rise to the quintessence-like behavior,
supporting the quintom-like behavior evidently.\footnote{In
\cite{obszx1}, the joint fitting of SNIa+CMB+LSS for the holographic
dark energy model gives the parameter constraints in 1 $\sigma$:
$c=0.81^{+0.23}_{-0.16}$, $\Omega_{\rm m0}=0.28\pm 0.03$, with
$\chi_{\rm min}^2=176.67$. In \cite{obszx2}, using the $f_{\rm gas}$
values provided by {\it Chandra} observational data (the X-ray gas
mass fraction of 26 rich clusters), the 1 $\sigma$ fit values for
$c$ and $\Omega_{\rm m0}$ are given: $c=0.61^{+0.45}_{-0.21}$ and
$\Omega_{\rm m0}=0.24^{+0.06}_{-0.05}$, with the best-fit chi-square
$\chi_{\rm min}^2=25.00$} Though the present result (in 1-$\sigma$
error range) from the analysis of the up-to-date observational data
does not support the quintom-like feature as strongly as before, the
best-fit value ($c=0.91$) still exhibits the holographic quintom
characteristic. However, the cases of $c<1$ will bring theoretical
problems: (i) This will lead to dark energy behaving as a phantom
eventually, which violates the weak energy condition of general
relativity, and the Gibbons-Hawking entropy will thus decrease since
the event horizon shrinks, which violates the second law of
thermodynamics as well. (ii) The quantum instability may often be
encountered in quintom models when the $w=-1$ crossing happens.
(iii) When the future event horizon as the IR cut-off becomes
shorter than the UV cut-off within a finite time in the future, the
definition of the holographic dark energy will break down.
Consequently, from the theoretical perspective, the holographic dark
energy with $c\geq 1$ is more reasonable. On the whole, since the
data analysis cannot rule out the possibility of $c \geq 1$
completely, the cases of $c \geq 1$ are worth investigating in
detail. In order to describe the holographic dark energy with the
the quintessence scalar-field (the low-energy effective theory), we
in this paper restrict $c\geq 1$ for the holographic dark energy
model.

According to the forms of quintessence energy density and pressure, one can
easily derive the scalar potential and kinetic energy term as
\begin{equation}
{\frac{V(\phi )}{\rho _{\mathrm{c0}}}}={\frac{1}{2}}(1-w_{\phi })\Omega
_{\phi }E^{2},  \label{potential}
\end{equation}%
\begin{equation}
{\frac{\dot{\phi}^{2}}{\rho _{\mathrm{c0}}}}=(1+w_{\phi })\Omega _{\phi
}E^{2},  \label{kinetic}
\end{equation}%
where $\rho _{\mathrm{c0}}=3M_{\mathrm{pl}}^{2}H_{0}^{2}$ is today's
critical density of the universe. Imposing the holographic nature (with $%
c\geq 1$) to the quintessence, the energy density of quintessence is needed
to satisfy the requirement of holographic principle, i.e., we should
identify $\rho _{\phi }$ with $\rho _{\Lambda }$. Then, the quintessence
field acquires the holographic nature, namely, $E$, $\Omega _{\phi }$ and $%
w_{\phi }$ are given by Eqs. (\ref{Ez}), (\ref{deq}) and (\ref{w}). Without
loss of generality, we assume $V^{\prime }>0$ and $\dot{\phi}<0$ in this
paper. Then, the derivative of the scalar field $\phi $ with respect to the
redshift $z$ can be given by
\begin{equation}
{\frac{\phi ^{\prime }}{M_{\mathrm{pl}}}}={\frac{\sqrt{3(1+w_{\phi })\Omega
_{\phi }}}{1+z}}.  \label{phiprime}
\end{equation}%
Consequently, we can easily obtain the evolutionary form of the field by
integrating the above equation
\begin{equation}
\phi (z)=\int\limits_{0}^{z}\phi ^{\prime }dz,  \label{phi}
\end{equation}%
where the field amplitude at the present epoch ($z=0$) is fixed to be zero,
namely $\phi (0)=0$. In what follows, we use the criterion $\phi (z)/M_{%
\mathrm{pl}}\leq 1$ to give the possible theoretical constraints on
the values of $c$ and $\Omega _{\mathrm{m0}}$.

%\begin{figure}[htbp]
%\centering
%\par
%\begin{center}
%$%
%\begin{array}{c@{\hspace{0.2in}}c}
%\multicolumn{1}{l}{\mbox{}} & \multicolumn{1}{l}{\mbox{}} \\
%\includegraphics[scale=0.65]{fig1a.eps} & %
%\includegraphics[scale=0.65]{fig1b.eps} \\
%\mbox{\bf (a)} & \mbox{\bf (b)}%
%\end{array}%
%$%
%\end{center}
%\caption{{\protect\small The cosmological evolution of the
%holographic dark energy. Panel (a) shows the evolution of
%$\Omega_{\Lambda}$, and panel (b) shows the evolution of
%$w_{\Lambda}$. Selected curves correspond to $c=2$, 3.5 and 5,
%respectively. Here the present fractional matter density is set to $\Omega_{%
%\mathrm{m0}}=0.28$.}}
%\label{fig1}
%\end{figure}

\begin{figure}[tbph]
\begin{center}
\includegraphics[scale=0.9]{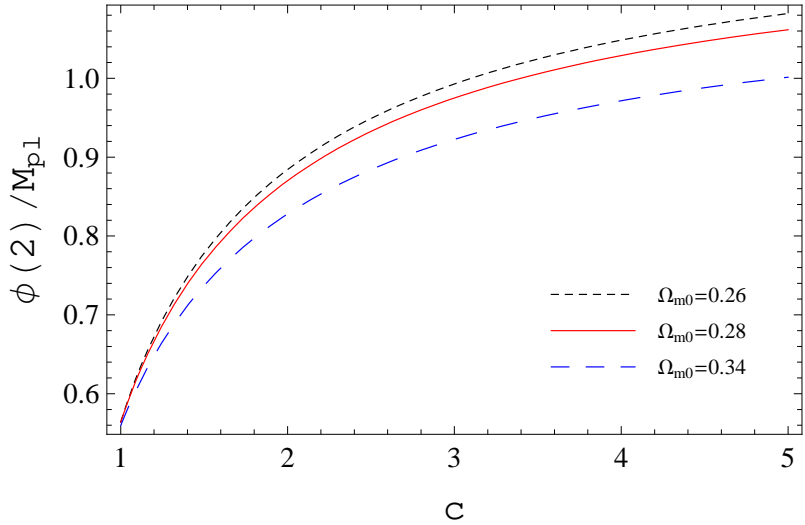}
\end{center}
\caption{{\protect\small The relationship between $\protect\phi(z)$(at $z=2$%
) and $c$. Selected curves correspond to
$\Omega_{\mathrm{m0}}=0.26$, 0.28 and 0.34, respectively.}}
\label{fig1}
\end{figure}

First, one can solve the equation (\ref{deq}) numerically and plot
$w_{\Lambda}(z)$ (see Fig. 1 in Ref. \cite{holoquin}). From that
figure, one can see that larger value of $c$ makes the value of
$w_{\Lambda}$ relatively larger. This makes the amplitude of field
$\phi (z)$ larger if the value of $c$ becomes larger (see Fig. 3 in
Ref. \cite{holoquin}). For making this point more clear, we plot
$\phi(z=2)$ versus $c$ in Fig. \ref{fig1}, where selected curves
correspond to $\Omega_{\mathrm{m0}}=0.26$, 0.28 and 0.34,
respectively. Thus, if $c$ becomes large, the value of $\phi$ will
become large (see Fig. \ref{fig1}) and in some cases, it may disobey
the criterion that the variation of quintessence scalar-field should
be less than $M_{\mathrm{pl}}$. Therefore, the criterion $\left\vert
\Delta \phi (z)\right\vert =\phi (z)\leq M_{\mathrm{pl}}$ is able to
give important theoretical constraints on the values of $c$ and
$\Omega_{\mathrm{m0}}$.

\begin{figure}[htbp]
\centering
\par
\begin{center}
$%
\begin{array}{c@{\hspace{0.2in}}c}
\multicolumn{1}{l}{\mbox{}} & \multicolumn{1}{l}{\mbox{}} \\
\includegraphics[scale=0.7]{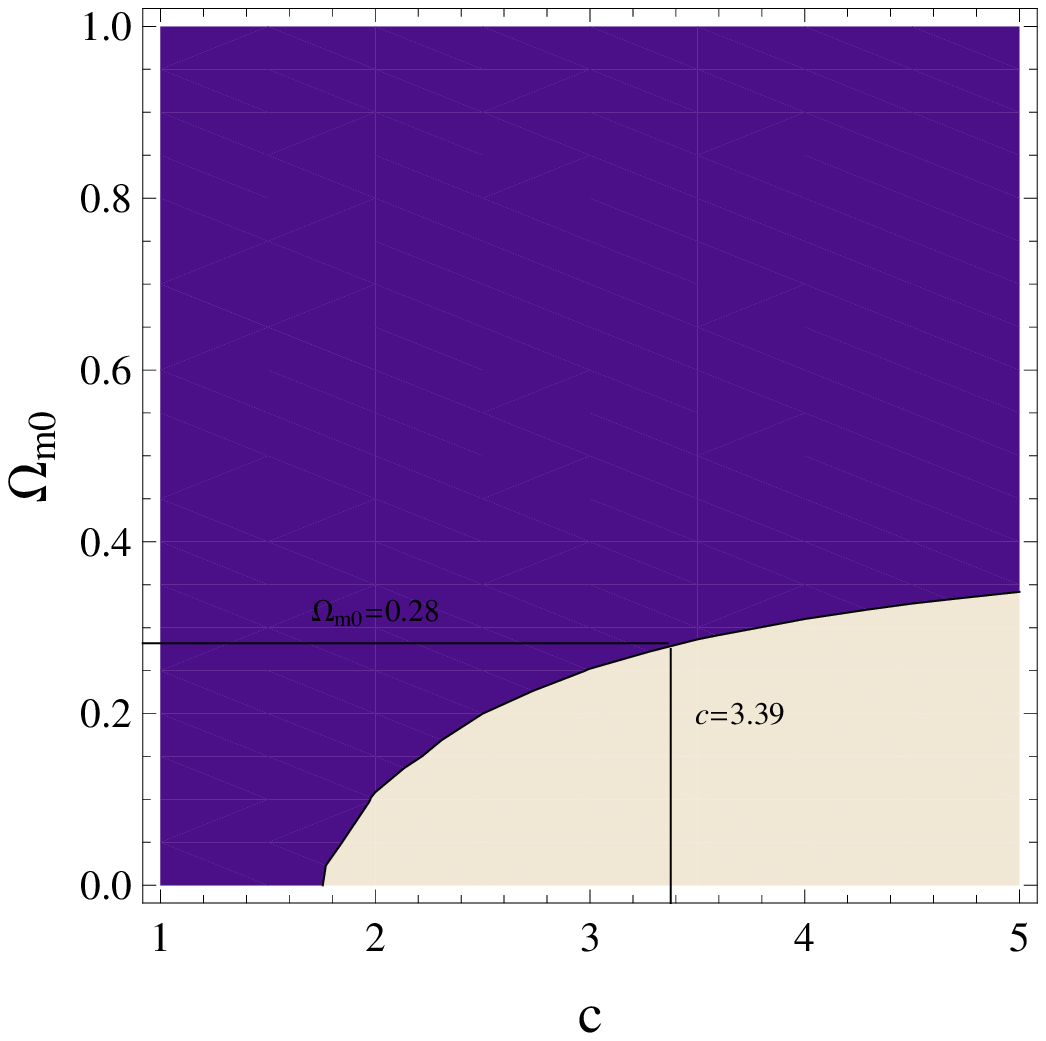} & %
\includegraphics[scale=0.85]{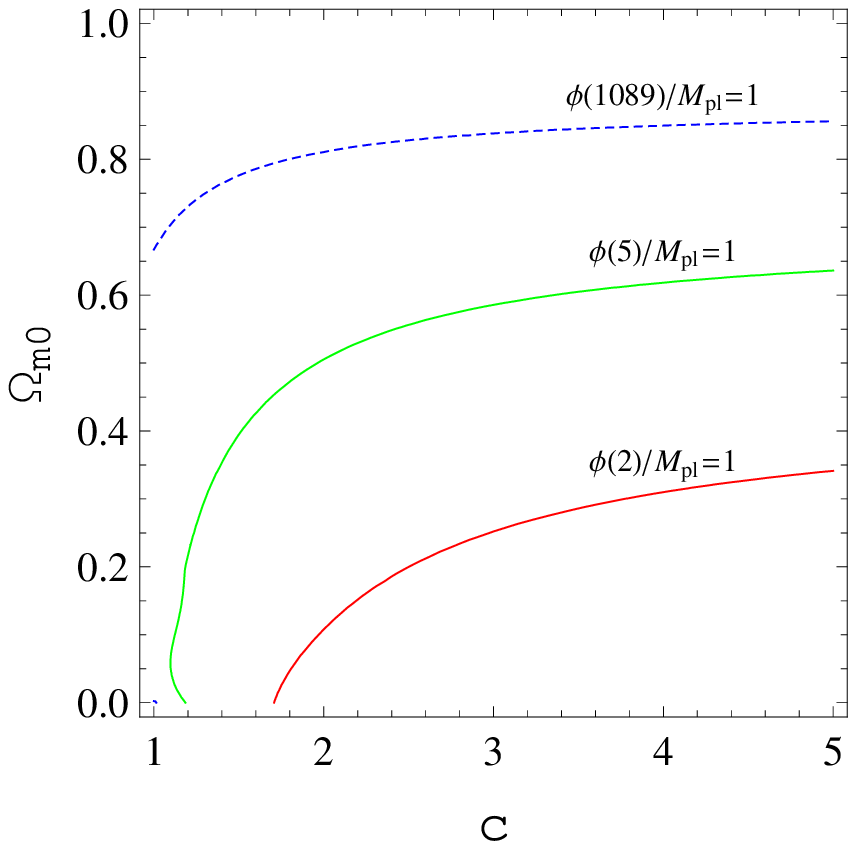} \\
\mbox{\bf (a)} & \mbox{\bf (b)}%
\end{array}%
$%
\end{center}
\caption{{\protect\small Panel (a): Constraint on the parameter-space of the
holographic quintessence from the theoretical criterion $\protect\phi(z=2)/M_{%
\mathrm{pl}}\leq 1$. Upper shaded area represents the allowed region. Panel
(b): Curves corresponding to $\protect\phi(z)/M_{\mathrm{pl}}=1$ in the $%
c-\Omega_{\mathrm{m0}}$ parameter-plane, where $z$ is taken to be 2,
5, and 1089, respectively.}} \label{fig2}
\end{figure}

\begin{figure}[tbph]
\begin{center}
\includegraphics[scale=0.9]{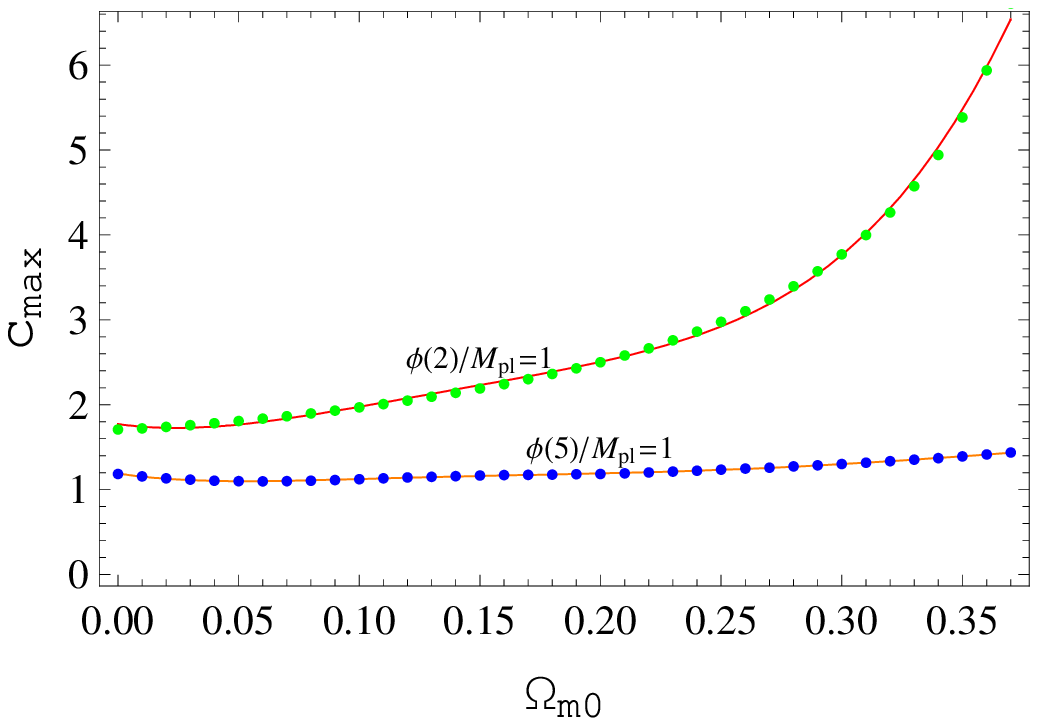}
\end{center}
\caption{\textrm{The empirical relations between $c_{max}$ and $\Omega_{m0}$%
. Points are generated from the equation $\protect\phi(z)/M_{\mathrm{pl}}=1$%
, where $z $ is taken to be 2 (green points) and 5 (blue points)
respectively, and curves are the numerical fitting results.}}
\label{fig3}
\end{figure}

Generically, the dark energy component is negligible in early times
of the universe. Hence, one should confirm when the dark energy
starts to operate in the universe. In general, the redshift at
$z\sim 2$ can be viewed as the onset of dark energy evolution, since
at which dark energy begins to take over the mantle of matter
component (albeit at that time matter component still dominates the
evolution of the universe). Therefore, we can set $z=2$ as the onset
of dark energy evolution. Of course, for the sake of safety, we can
also take, say, $z=5$ as the onset of dark energy evolution. An
example is shown in the panel (a) of Fig. {\ref{fig2}, where $z=2$
is taken. This figure shows the constraints for the $c-\Omega
_{\mathrm{m0}}$ parameter-space of the
holographic quintessence from the theoretical criterion $\phi (z)/M_{\mathrm{%
pl}}\leq 1$, where the borderline is set by $\phi
(z=2)/M_{\mathrm{pl}}=1$, and the allowed region of the
parameter-space is represented by the shaded area. For comparison,
we also plot the curves corresponding to $\phi
(z=5)=M_{\mathrm{pl}}$ and $\phi (z=1089)=M_{\mathrm{pl}}$ in the
$c-\Omega _{\mathrm{m0}}$ plane in the panel (b) of Fig. \ref{fig2},
which shows that the borderline will get an upper shift when
enlarging the redshift, leading to the allowed region shrinks. Note
that the case of $z=1089$ is not an appropriate example because at
the time of CMB formation the universe is dominated by
non-relativistic matter and the dark energy is totally negligible. }

The theoretical limit of $c$ is shown explicitly in the panel (a) of Fig. %
\ref{fig2}. When fixing the value of $\Omega _{\mathrm{m0}}$, the
upper
bound of $c$ can be read from this figure directly. For example, choosing $%
\Omega _{\mathrm{m0}}=0.28$ which is favored by the current observations,
the criterion $\phi (2)/M_{\mathrm{pl}}\leq 1$ directly leads to $c_{\mathrm{%
max}}=3.39$ which is the theoretical limit of the parameter $c$. However,
since the relationship between $c_{\mathrm{max}}$ and $\Omega _{\mathrm{m0}}$
is derived from Eq. (\ref{phi}) which is an integral formula making the
relation $c_{\mathrm{max}}(\Omega _{\mathrm{m0}})$ difficult to identify, we
should furthermore find an empirical relation between $c_{\mathrm{max}}$ and
$\Omega _{\mathrm{m0}}$. Thus, we output the data along the curve $\phi
(z=2)/M_{\mathrm{pl}}=1$ and fit them with the elementary functions, then we
obtain
\begin{equation}
c_{\max }(\Omega _{\mathrm{m0}})=15341.8~e^{\Omega
_{m0}}-15340.1-15342.9~\Omega _{\mathrm{m0}}-7609.9~\Omega
_{m0}^{2}-2875.4~\Omega _{\mathrm{m0}}^{3},  \notag
\end{equation}%
which is very easy for us to operate. For instance, when substituting $%
\Omega _{\mathrm{m0}}=0.24$ in it, it gives $c_{\mathrm{max}}=2.77$; when
substituting $\Omega _{\mathrm{m0}}=0.28$, it gives $c_{\max }=3.34$.
Therefore, this empirical function $c_{\max }(\Omega _{\mathrm{m0}})$ is
very convenient for us to get the theoretical limit of the parameter $c$.
Likewise, we can also get an empirical relation $c_{\max }(\Omega _{\mathrm{%
m0}})$ for the $z=5$ case. We do not exhibit this case explicitly
here, but gives an output point as example: the input $\Omega
_{\mathrm{m0}}=0.28$ gives $c_{\mathrm{max}}=1.27$. The numerical
fitting curves with data points are shown in Fig. \ref{fig3}.

To summarize, in this paper we investigate the possible theoretical
limits on the parameter $c$ of the holographic quintessence. We
adopt the perspective that the scalar-field model is an effective
description for the underlying theory of dark energy. In the
holographic dark energy model, the equation of state with $c\geq 1$
evolves within the range $-1\leq w\leq -1/3$, so it looks like a
quintessence. Quintessence scalar-field can thus be used to
effectively describe the holographic dark energy with $c\geq 1$. For
quintessence scalar-field, the requirement (from the weak gravity
conjecture) that the variation of the field should be less than
$M_{\mathrm{pl}}$ will set a theoretical bound on the model. So, in
this paper, we tested the holographic quintessence model using this
criterion and obtained the theoretical limits on the parameter $c$
for the model. Anyway, the theoretical limits discussed in this
paper is only a possibility. The requirement that the variation of
the canonical quintessence field minimally coupled to gravity is
less than the Planck scale may arise from the consistent theory of
quantum gravity. In this sense, the results derived in this paper
can, to some extent, be viewed as the prediction of quantum gravity.
Though the constraints on the parameter $c$ are rather loose, the
possible theoretical limits of the holographic quintessence model
are worth investigating.

\section*{Acknowledgements}

One of us (YZM) would like to thank Miao Li, Qing-Guo Huang,
Rong-Gen Cai and Yan Gong for helpful discussions. He also thanks
Jian Ma, Nan Zhao and Hao Yin for helpful direction of computer
program. This work was supported partially by grants from NSFC (No.
10325525, No. 90403029, No. 10525060 and No. 10705041) and grants
from the China Postdoctoral Science Foundation, the K. C. Wong
Education Foundation and the Chinese Academy of Sciences.

\end{document}